\documentstyle[twocolumn,aps,epsfig,graphics]{revtex}

\begin{document}

\title{Production of $\Theta^+(1540)$ and $\Xi$ Pentaquark States in 
Proton-Proton Interactions}

\author{M.~Bleicher{}$^a$, F.M.~Liu{}$^{a,b,}$\thanks{Alexander von Humboldt Fellow}, 
J.~Aichelin{}$^c$, 
T.~Pierog{}$^d$, K.~Werner{}$^c$\vspace*{.5cm}}

\address{{}$^a$ Institut f\"ur Theoretische Physik, J. W. Goethe Universit\"at, \\
Robert-Mayer-Str. 10, 60054 Frankfurt am Main, Germany}

\address{{}$^b$ Institute of Particle Physics , Huazhong
Normal University, Wuhan, China}

\address{{}$^c$ SUBATECH, Laboratoire de Physique Subatomique et des Technologies Associ\'{e}es
\\
 University of Nantes - IN2P3/CNRS - Ecole des Mines de Nantes \\
 4 rue Alfred Kastler, F-44072 Nantes Cedex 03, France}

\address{{}$^d$ FZK, Institut f\"ur Kernphysik, Karlsruhe, Germany}

\maketitle
The production of strange pentaquark states (e.g. Theta baryons and $\Xi^{--}$ states) 
in hadronic interactions
within a Gribov-Regge approach is explored. In this approach the \( \Theta^+(1540) \) and
the $\Xi$
are produced by disintegration of remnants formed by the exchange of pomerons between 
the two protons.
We predict the rapidity and transverse momentum distributions as well as the $4\pi$ 
multiplicity  of the \( \Theta^+ \), $\Xi^{--}$, $\Xi^-$, $\Xi^0$ and $\Xi^+$ 
for $\sqrt{s}$ = 17~GeV (SPS) and 200~GeV (RHIC). 
For both energies more than $10^{-3}$ \( \Theta^+ \) and more than $10^{-5}$ $\Xi$ 
per pp event should be observed by the present experiments. 

\vspace{.6cm}
Very recently in photon-nucleus \cite{Nakano:2003qx,Stepanyan:2003qr} 
and Kaon-nucleus experiments \cite{Barmin:2003vv} a new baryon, consisting of
five quarks, uudd$\overline{\rm s}$,  has been identified in the $K^+n$ or $K^0p$ 
invariant mass spectrum. It has been been named $\Theta^+$ particle and has 
spin 1/2, isospin 0 and strangeness +1. Its mass is about 1.54 GeV and its width
is less than 25 MeV. Such a state has been predicted by 
 Diakonov \cite{Diakonov:1997mm} in the framework of a chiral soliton model.

This finding has renewed the experimental and theoretical interest for
novel baryon states. Major progress has been reported on a possible extension of
the original model \cite{Diakonov:2003ei,Diakonov:2003jj}, 
within the Skyrme model \cite{Praszalowicz:2003ik,Itzhaki:2003nr,Borisyuk:2003fk}, 
and within the constituent quark model \cite{Jaffe:2003sg,Stancu:2003if}.
Also (lattice) QCD studies of 
the $\Theta^+$ (see e.g. \cite{Csikor:2003ng,Sugiyama:2003zk,Sasaki:2003gi}) have 
been performed and first explorations of the $\Theta^+$ multiplicity at SPS and 
RHIC energies
are available \cite{Randrup:2003fq,Liu:2003rh,Letessier:2003by}.
Thus, the existence of this novel state has many perspectives 
for pp as well as for nucleus-nucleus collisions. 

In this letter, we present predictions for the $\Theta^+$ in pp collisions 
from a newly developed approach for hadronic interactions. 
It has recently been shown \cite{Bleicher:2001nz} that the standard string 
fragmentation models which described spectra and 
multiplicities of many hadrons rather well need to be revised: due to their 
diquark-quark topology these models produce more $\overline \Omega$ 
than $\Omega $ in medium and low energetic pp interactions, in contradistinction to
experiments. Therefore a key issue is presently to gain information on the details of
hadron production. Especially more exotic states like the $\Lambda (1405)$ which
may be a udsu$\overline{\rm u}$ state or the $\Theta^+$ carry important information
to tackle this question. If
their multiplicity can be related to that of other particles one can hope to get
experimentally a handle on the hadronization process.

In heavy ion collisions the multiplicity of the most abundant particles 
can be well described in a statistical model assuming a temperature
close to that where the chiral/confinement phase transition is expected 
and a moderate chemical potential. Unstable particles can test how the
expanding system interacts afterwards because if the decay products have still
an interaction in the invariant mass spectra the resonance cannot be 
identified anymore. Especially long living states are very useful in this
respect. 

Until this year, the search for multi-quark bags focused mainly 
on the  $H$-particle \cite{Jaffe:1976yi},
because it is closely related to the study of $\Xi$ 
and $\Lambda\Lambda$ hypernuclei (see e.g.  \cite{Ahn:sx,Takahashi:nm}). 
The $H$ is a six quark state (uuddss) coupled to an SU(3) singlet in color and flavour.
Unfortunately no stringent observation of the $H$-particle exists. Even today, decades after the 
first prediction of the $H$-di-baryon by 
Jaffe \cite{Jaffe:1976yi} the question of its existence is still open.

In contrast to the $H$ particle, the situation for the $\Theta^+$ baryon 
is very promising. Thus, in this letter we explore the formation of 
the $\Theta^+$-baryon within a new
approach called parton-based Gribov-Regge theory. It is realized in the 
Monte Carlo  program {\small NE}{\large X}{\small US} 3.97 \cite{Drescher:2000ha,Liu:2002gw}. 
In this model  high energy hadronic
and nuclear collisions are treated within a self-consistent quantum
mechanical  multiple scattering formalism. Elementary interactions,
happening in  parallel, correspond to underlying microscopic (predominantly 
soft) parton  cascades and are described effectively as phenomenological
soft Pomeron exchanges.
A Pomeron can be seen as layers of a (soft) parton ladder,
which is attached to projectile and target nucleons via leg partons. 
At high energies one accounts also for the contribution of perturbative 
(high $p_t$) partons described by a so-called "semihard Pomeron" - a piece 
of the QCD parton ladder sandwiched between two soft Pomerons which are 
connected to the projectile and to the target in the usual way. The spectator 
partons of both projectile and target nucleons, left after Pomeron emissions, 
form nucleon remnants. 
The legs of the Pomerons form color singlets, such as q-\(\overline{\mathrm{q}} \),
q-qq or \( \overline{\mathrm{q}} \)-\( \overline{\mathrm{q}} \)\( \overline{\mathrm{q}} \).
The probability of q-qq and \( \overline{\mathrm{q}} \)-
\( \overline{\mathrm{q}} \)\( \overline{\mathrm{q}} \)
is controlled by the parameter \( P_{\mathrm{qq}} \) and is 
fixed by the experimental yields on (multi-) strange baryons \cite{Liu:2002gw}. 
\begin{figure}
 \par\resizebox*{!}{0.11\textheight}{\includegraphics{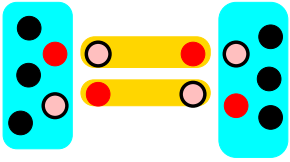}} \par{}
\vspace*{.8cm}
\caption{The typical collision configuration has two remnants and one cut Pomeron
represented by two \protect\protect\( \mathrm{q}-\overline{\mathrm{q}}\protect \protect \)
strings.  In the case of a uudd$\overline{\rm s}$ 
flavour content a \protect\protect\( \Theta^+\protect \protect \) baryon can form. 
\label{quarkbag}}
\end{figure}

Particles are then produced from cutting the Pomerons and the decay of the remnants. 
As an intuitive way to understand particle production, each cut Pomeron 
is  regarded as two strings, i.e. two layers of a parton ladder.   
Each string has two ends which are quark(s) or antiquark(s) from the two 
Pomeron legs respectively.
To compensate the flavour, whenever a quark or an antiquark is taken as 
a string end, a corresponding anti-particle is put in the remnant nearby.

Since an arbitrary number of Pomerons may be involved, it is natural to
take quarks and antiquarks from the sea as the string ends.   
In order to describe the experimental yields on (multi-)strange baryons 
\cite{Liu:2002gw}, all the valence quarks stay in the remnants, 
whereas the string ends are represented by sea quarks.
Thus, Pomerons are vacuum excitations and produce particles and 
anti-particles equally.
Note that in addition to these singlet type processes, valence quark hard
interactions are treated differently in the present model.
To give a proper description of deep inelastic scattering data, a certain 
fraction of the Pomerons is
connected to the valance quarks of the hadron, not leading 
to a quark feeding of the remnant. 
This kind of hard processes will not be discussed here, but is included in 
the simulation. 
Only the remnants change the balance of particles and anti-particles,
due to the valence quarks inside resulting in the possibility to solve 
the anti-omega puzzle 
\cite{Bleicher:2001nz} at the SPS.

This prescription is able to accumulate multiple quarks and anti-quarks in the remnants
depending on the number of exchanged Pomerons. In the most simple case of a
single Pomeron exchange, the remnant may gain an additional anti-quark and a
quark and is transformed into a pentaquark bag as discussed in the following.

The typical collision configuration has two remnants and one cut Pomeron represented
by two \protect\( \mathrm{q}-\overline{\mathrm{q}}\protect  \) strings, see
Fig. \ref{quarkbag}.
Four quarks and one anti-quark are now in the remnant: the
three valence quarks  u, u, d  plus a q$_i$$\overline{\rm q}_j$ sea quark pair.
Each of the q$_i$ may have the flavour  u , d, or  s, with relative weights
1 : 1 : \( f_{\mathrm{s}} \). The parameter \( f_{\mathrm{s}} \), is
fixed by  strange hadron data in proton-proton scatterings at
160~GeV to be \( f_{\mathrm{s}}=0.26 \). Thus, there is a
small but nonzero probability to have a uudd$\overline{\rm s}$ flavour in a 
remnant, such that a \( \Theta^+ \) pentaquark may be formed. 
\begin{figure}
\vspace*{-.5cm}
\par \resizebox*{!}{0.35\textheight}{\includegraphics{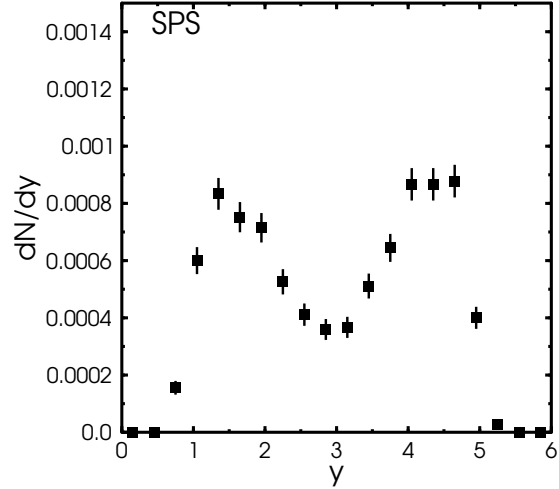}} \par{}
\caption{Rapidity distributions of \protect\protect\( \Theta^+ \protect \protect \)'s in
pp interactions at E\protect\( _{\textrm{lab}}=160\protect \)~GeV. 
\label{dndy}}
\end{figure}
\begin{figure}
\vspace*{-.5cm}
\par \resizebox*{!}{0.35\textheight}{\includegraphics{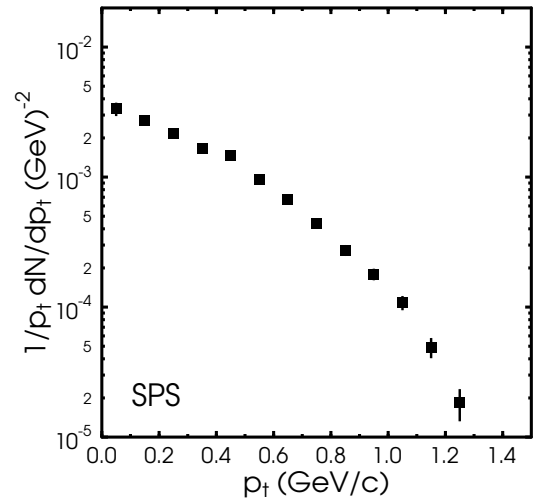}} \par{}
\caption{Transverse momentum distributions of \protect\protect\( \Theta^+\protect \protect \)'s
in pp interactions at E\protect\( _{\textrm{lab}}=160\protect \)~GeV. 
\label{dndpt}}
\end{figure}

The remnants have mass 
distribution \( P(m^{2})\propto (m^{2})^{-\alpha },\, \, \, m^{2}\in 
(m^{2}_{\mathrm{min}},\, x^{+}s), \)
here \( s \) is the squared CMS energy. With, \( m_{\mathrm{min}} \) being the minimal
hadron mass compatible with the remnant's quark content, and \( x^{+} \) is
the light-cone momentum fraction of the remnant which is determined in the collision
configuration. In the present study, the parameter \( \alpha  \) is 2.25 for non diffractive
interactions and 1 for diffractive events\cite{kw1}.
This remnant disintegrates into hadrons according to (microcanonical) 
phase space \cite{Werner:1995mx}. 
This approach
describes quite well multiplicity\footnote{A variation of the model
parameters by 20\% results in variation of the model results by roughly 10\%
both for the Pentaquarks and the well known hadrons. A larger variation of the parameters
destroys the agreement with the NA49 pp data at 160 GeV.}, transverse momenta and rapidity of all the observed hadrons
\cite{kw1} in pp collisions at energies between 40 and 160 GeV.

For the present study we have embedded the \( \Theta^+ \) in the microcanonical 
approach to describe the
disintegration of the remnant. It is therefore treated the same way as all the other hadrons. 
A similar approach was recently introduced to study $H^0$ production in pp 
interactions\cite{Bleicher:2002cy}.

As discussed above, a pp collision in the present model involves two
sources of particles production:
\begin{enumerate}
\item Fragmentation of strings (from exchanged Pomerons).
\item Decay of the projectile/target remnant.
  The decay of the remnant can happen in two distinct ways:
 \begin{itemize}
 \item[2a.)]by string fragmentation or
 \item[2b.)]by micro-canonical phase space.
  \end{itemize}
\end{enumerate}

Presently, Pentaquarks can only be produced from the micro-canonical
remnant decay. A fragmentation of strings into Pentaquarks is not
included.

This is in contrast to other hadrons, e.g. Lambdas that can be produced 
from all mentioned sources.

There is no difference in the general input parameters for calculations
for $\Lambda$s and Pentaquarks. The main parameter is the probability for
having a (anti-)strange quark at the Pomeron leg. This parameter controls
strangess (e.g. Lambda and Kaon multiplicities) and Pentaquark production directly.

The additional inputs for the study of Pentaquark states are
their masses, quark contents and spin, which have ultimately to be
determined by the experimental results. Presently, we use the values
from Ref. \cite{Diakonov:1997mm,Diakonov:2003ei,Diakonov:2003jj}.

Contrary to grand canonical thermal model estimates \cite{Randrup:2003fq,Letessier:2003by}, 
which implicitly assume chemical and thermal equilibrium, 
the present approach works differently:
In the present approach, initial and final state 
dynamics of the pp interaction are fully taken into account, thus allowing to 
predict transverse and longitudinal
momentum distributions all produced hadrons, including the $\Theta^+$.
\begin{figure}
\vspace*{-.5cm}
\par \resizebox*{!}{0.35\textheight}{\includegraphics{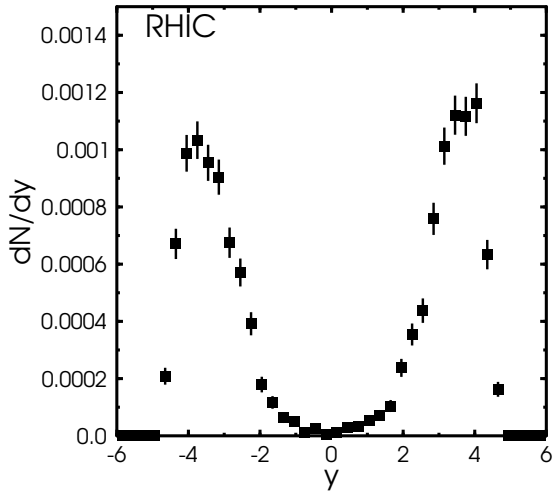}} \par{}
\caption{Rapidity distributions of \protect\protect\( \Theta^+\protect \protect \)'s in
pp interactions at \protect\( \sqrt{s}=200\protect \)~GeV. 
\label{dndy200}}
\end{figure}
\begin{figure}
\vspace*{-0.5cm}
\par \resizebox*{!}{0.35\textheight}{\includegraphics{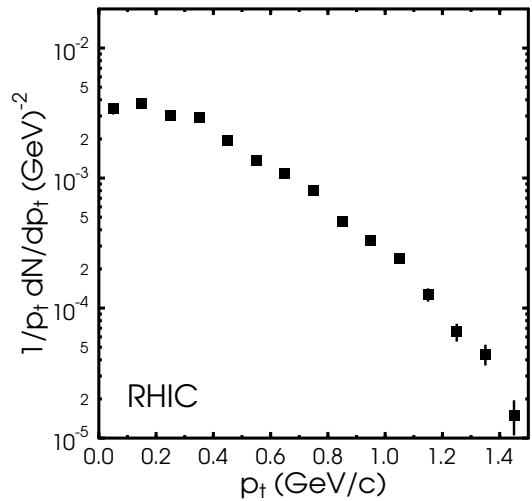}} \par{}
\caption{Transverse momentum distributions of \protect\protect\( \Theta^+\protect \protect \)'s
in pp interactions at \protect\( \sqrt{s}=200\protect \)~GeV. 
\label{dndpt200}}
\end{figure}
\begin{table}
\begin{tabular}{lcrr}
Particle  & Quark content          & Yield (SPS)        & Yield (RHIC)\\\hline
$\Theta^+$&(uudd$\overline{\rm s}$)&$3.5\times 10^{-3}$ &$5.2\times 10^{-3}$ \\
$\Xi^+   $&(ssuu$\overline{\rm d}$)&$4.8\times 10^{-5}$ &$11.9\times 10^{-5}$ \\
$\Xi^0   $&(ssud$\overline{\rm d}$)&$4.7\times 10^{-5}$ &$9.2\times 10^{-5}$ \\
$\Xi^{-} $&(ssud$\overline{\rm u}$)&$3.0\times 10^{-5}$ &$6.3\times 10^{-5}$ \\
$\Xi^{--}$&(ssdd$\overline{\rm u}$)&$1.8\times 10^{-5}$ &$4.5\times 10^{-5}$ \\
\end{tabular}
\caption{\label{table1} Predictions of the various Pentaquark abundances in $4\pi$
for inelastic pp collisions at $E_{\rm lab}=160$~GeV (SPS) and $\sqrt s = 200$~GeV (RHIC).}
\end{table}

\noindent
Furthermore, a microcanonical model is employed for the remnant decay which counts exactly 
the phase space states for a given energy 
and volume of the system. Only in the limit of large volumes and many particles it agrees
with the  grand canonical approaches. 

Let us now study the multiplicities and momentum spectra of the calculated \( \Theta^+ \)'s.
Fig. \ref{dndy} depicts the rapidity distribution of the predicted \( \Theta^+ \)'s at the 
top SPS energy. One observes a slight dip around central rapidities due to the fact that the
particles originate from remnants.
The total $\Theta^+$ multiplicity per inelastic event is 0.0035.  
Table \ref{table1} gives also the predicted yields of the
other species of Pentaquarks. It is interesting to note that
the $\Xi$ Pentaquark is suppressed by two orders of magnitude compared to the Theta
particle. This is due to the double strange quark component of the $\Xi$ Pentaquark
that is suppressed at Pomeron legs. 
However, it is interesting to note that  the double-strange 
$\Xi$ Pentaquark has been well observed at this energy by the NA49
collaboration. Thus, with the presently accumulated statistics of the NA49 
collaboration, the \( \Theta^+ \) should be easily visible in the data.
Fig. \ref{dndpt} depicts the transverse momentum spectra of the \( \Theta^+ \)'s.
In Figures \ref{dndy200} and \ref{dndpt200} we show the rapidity and transverse
momentum spectra at \( \sqrt{s}=200 \)~GeV. 
Especially at RHIC energies one clearly observes the pile-up of
$\Theta^+$ baryons in the forward and backward hemisphere. In the midrapidity region
the pentaquark yield vanishes. 
It should be noted that the vanishing yield
at central rapidities might be filled up, if also the (yet unknown) 
fragmentation function into Pentaquarks in a string break up is included.

In conclusion, we have studied the production of Pentaquark baryons 
in pp collisions from parton-based Gribov-Regge theory. 
Multiplicities, rapidity and transverse
momentum spectra for the $\Theta^+$ and $\Xi$ Pentaquarks are predicted for pp 
interaction at E\( _{\textrm{lab}}=160 \)~GeV and \( \sqrt{s}=200 \)~GeV. 
We predict a total yield of $3-5 \times 10^{-3}$ $\Theta^+$ per inelastic pp 
event at SPS and RHIC. The $\Xi$ pentaquark states are suppressed by two orders
of magnitude compared to the $\Theta^+$. 
Thus, our predictions are  accessible  
by the NA49 experiment at CERN and the STAR experiment at RHIC.

\section{Acknowledgements}

This work was supported by GSI, DFG, BMBF.
FML thanks the Alexander von Humboldt foundation for financial support.

\end{document}